\documentclass[superscriptaddress,aps,prd,showpacs,preprint]{revtex4-1}

\usepackage{eurosym}
\usepackage{graphicx}
\usepackage{amsmath}
\usepackage{mathrsfs}
\usepackage{bm}
\usepackage{color}
\usepackage{amsmath}
\usepackage{amssymb}

\usepackage{multirow}
\usepackage{array}

\usepackage{booktabs, makecell, longtable}

\def\be{\begin{equation}}
\def\ee{\end{equation}}
\def\bea{\begin{eqnarray}}
\def\eea{\end{eqnarray}}

\allowdisplaybreaks

\begin{document}
\title{Near-horizon quasinormal modes of charged scalar around a general spherically symmetric black hole}

\author{Takol Tangphati}
\email{takoltang@gmail.com}
\affiliation{High Energy Physics Theory Group, Department of Physics,
	Faculty of Science, Chulalongkorn University, Phyathai Rd., Bangkok 10330, Thailand}
\author{Supakchai Ponglertsakul}
\email{supakchai.p@gmail.com}
\affiliation{High Energy Physics Theory Group, Department of Physics,
	Faculty of Science, Chulalongkorn University, Phyathai Rd., Bangkok 10330, Thailand}
\author{Piyabut Burikham}
\email{piyabut@gmail.com}
\affiliation{High Energy Physics Theory Group, Department of Physics,
	Faculty of Science, Chulalongkorn University, Phyathai Rd., Bangkok 10330, Thailand}
	
\date{\today }

\begin{abstract}
We study the quasinormal modes (QNMs) of charged scalar in the static spherically symmetric black hole background near the event and cosmological horizon. Starting with numerical analysis of the QNMs of black hole in the dRGT massive gravity, the mathematical tool called the Asymptotic Iteration Method~(AIM) is used to calculate the quasinormal frequencies. The parameters such as the mass and charge of the black hole, the cosmological constant, the coefficient of the linear term from massive gravity $\gamma$, and the mass of the scalar are varied to study the behavior of the QNMs. We found the tower pattern of the near-horizon quasinormal frequencies from the numerical results by AIM where the real parts depend only on product of the charge of the black hole and the scalar field and the imaginary parts depend only on the surface gravity. To confirm the numerical finding, we analytically determine the exact QNMs of the charged scalar near the horizons of any static spherically symmetric black hole background in the simple universal forms; $\omega = \displaystyle{\frac{qQ}{r_h}} + i \kappa_h n$ and $\omega = \displaystyle{\frac{qQ}{r_c}} + i |\kappa_c| n$ where $n$ is a non-positive integer and $\kappa_{h,c}$ is the surface gravity, for the event and cosmological horizon respectively.  Extending our analysis, we also compute the four towers of the near-horizon QNMs that can reach the far region.  The four kinds of QNMs converge to certain asymptotic values with equally spacing imaginary parts and the real parts proportional to $qQ/r_{h,c}$.  These modes do not match with the all-region~(WKB) modes of the real background since they are originated from the linearly approximated metric.
	
\vspace{1mm}
	
	
\end{abstract}

\maketitle

\section{Introduction}

Black hole has a unique property as its boundary is defined by the event horizon from within which nothing even light can escape.  Dynamics of the spacetime around the black hole is governed by the Einstein field equations, the event horion can be perturbed and the fluctuations of the spacetime can propagate and carry energy away from the black hole in the form of gravitational waves.  Oscillation modes of horizon usually have small amplitudes due to the high scale of the Planck mass which results in extreme stiffness of spacetime.  However, the frequencies of the oscillating horizon as well as any fields around it are determined by the black hole physical parameters such as mass, charge and spin.  In 2015, LIGO remarkably detected the gravitational waves from the merging of two black holes~\cite{Abbott:2016blz}.  The ringdown frequency profile of the black hole after binary merging is characterized by the quasinormal modes~(QNMs) of the black hole.      

QNM of a black hole is analogous to the damped oscillation of an oscillator.  The real part of QNM represents the energy or physical oscillating frequencies whilst the imaginary part gives the decaying time or relaxation time, it represents dissipation of the energy.  QNMs of a black hole are thus signature of each black hole that provides information of the intrinsic physical parameters, i.e., mass, charge and spin according to the General Relativity~(GR).  A vast number of studies of black hole QNMs were explored in many contexts. We refer interesting readers to nice reviews on the subject in Ref.~\cite{Kokkotas:1999bd,Berti:2009kk,Konoplya:2011qq}.

GR is the theory describing gravitational interaction mediated by a massless spin-2 particle called graviton~(in its quantized extension which is still problematic). Many phenomena had been predicted by GR and were experimentally confirmed e.g. gravitational time dilation \cite{Uggerhoj:2016,Schwartz:1907I}, precession of Mercury's orbit \cite{Clemence:1947} and gravitational waves \cite{Abbott:2016blz}. GR works incredibly well in many scenarios, however it fails to give a satisfying explanation to an accelerated expansion of the universe \cite{Riess:1998cb,Perlmutter:1998np} in the extragalactic scale and the asymptotically flat rotation curves of the galaxies~\cite{Konno:2008np}. This failure indicates that either there are overwhelmingly unknown form of dark matter and dark energy in the Universe or GR itself may need to be modified at such scales.  In contrast to GR, one can consider the theory where gravity is mediated by a massive graviton. This type of alternative theory of gravity is formally known as massive gravity.

The theory of massive gravity was originated long before the discovery of accelerated expansion of the universe. By adding an appropriated mass term (of graviton) to the linearised Einstein's general relativity, Fierz and Pauli firstly constructed a linear theory of massive gravity in 1939 \cite{Fierz:1939ix}. However, the mass term in Fierz-Pauli (FP) massive gravity leads to van Dam-Veltman-Zakharov (vDVZ) discontinuity \cite{Zakharov:1970cc,vanDam:1970vg} where the prediction of linear theory of massive gravity in massless limit does not agree with those of GR. It was realized later that the vDVZ discontinuity was in fact caused by the extra degree of freedom \cite{Vainshtein:1972sx}. Vainshtein suggested \cite{Vainshtein:1972sx} that the vDVZ discontinuity can be eliminated by considering the massive theory in the non-linear regime. However the generic non-linear theory of massive gravity usually suffers from the Boulware-Deser (BD) ghost \cite{Hinterbichler:2011tt}. Nevertheless, de Rham, Gabadadze, and Tolley (dRGT) have successfully constructed the non-linear theory of massive gravity without the BD ghost in 2010 \cite{deRham:2010kj}. Since then the dRGT has become useful in cosmological study. Its cosmological solution can be used to explain an accelerated expansion of the universe in the sense that the existence of graviton mass naturally provides the effect of cosmological constant \cite{Gumrukcuoglu:2011ew,Gumrukcuoglu:2011zh}. Moreover, the dRGT model can also explain a vast varieties of the galactic rotation curves~\cite{Panpanich:2018cxo}.

On the other hand, the physics of compact object in dRGT massive gravity is also interesting. A static black hole solution in $(n+2)$ dimensional massive gravity is found in \cite{Cai:2014znn}. Static spherically symmetric black holes with and without electric charge in dRGT massive gravity are constructed and investigated in \cite{Ghosh:2015cva}. Similar to cosmological solution, the effective cosmological constant appears naturally. Therefore one can consider these black holes as generalized version of Schwarzschild/Reissner-Nordstr\"om black holes with positive and negative cosmological constant.  In this work, we shall focus particularly on the QNMs of black holes in dRGT massive gravity \cite{Kokkotas:1999bd,Berti:2009kk,Konoplya:2011qq} as the generalized form of the black hole spacetime. 


In \cite{Burikham:2017gdm}, numerical study of the QNMs of charged dRGT black hole reveals some intriguing behaviour about the real parts and imaginary parts of the QNMs of charged scalar near the event and cosmological horizon. The real parts can be explicitly shown to be proportional to the product of the charge of scalar field and the black hole. While the imaginary parts exhibit some relation with the surface gravity at each horizon. Here in this article we try to fill the gap by providing an analytical explanation to the observed behaviour for those near horizon modes found in \cite{Burikham:2017gdm} and make corrections to the numerical procedure and results of Ref.~\cite{Burikham:2017gdm}. This paper is organized as follow. In Sec.~\ref{secset}, we setup the equation of motion for the charged scalar field in the curved black hole background and review the asymptotic iteration method (AIM)~\cite{AIM:2003,Cho:2009cj,Cho:2011sf,Prasia:2016fcc} which shall be employed to the numerical study of the QNMs in dRGT massive gravity. Then we highlight some intriguing behaviour of our numerical results in Sec.~\ref{secNUM}. In Sec.~\ref{sec_exact}, we provide analytical calculations of the QNMs for near event horizon and near cosmic horizon modes for any static and spherically symmetric black hole background. Section \ref{secDiscussion} devotes to discussions and conclusions. In this article, natural units $c = 1, \hbar = 1, \text{ and } G = 1$ are used throughout.

\section{The Setup and Numerical Analysis}\label{secset}

We consider perturbation of a charged scalar particle around the black hole spacetime. The dynamic of the field is described by the generalized Klein-Gordon equation with the presence of the gauge field,
\begin{equation}
(\nabla_{\alpha} - i q A_{\alpha})(\nabla^{\alpha} - i q A^{\alpha})\phi - m_s^2\Phi = 0,
\label{KG_general}
\end{equation}
where $\Phi$ is the scalar field, $\nabla_{\alpha}$ is the covariant derivative, $q$ is the charge of the scalar field, $A_{\mu}$ is the gauge field, and $m_s$ is the scalar field's mass.  The static spherically symmetric black hole metric can be expressed as
\begin{equation}
ds^2 = -f(r)dt^2 + \frac{dr^2}{f(r)} + r^2 d\Omega^2.
\label{metric_gen}
\end{equation}

First we will use the iterative method called the Asymptotic Iteration Method (AIM) in order to find the numerical results of quasinormal frequencies of the field in the generic black hole spacetime with mass, charge, linear term in radial coordinate $r$ and the cosmological constant term. A number of exotic matters and modified gravity models can give such background, a well-known example is the dRGT massive gravity~\cite{Ghosh:2015cva,Panpanich:2018cxo}. The process of AIM is split into two steps. The first one is to change the coordinates of the Klein-Gordon equation and to scale out the divergences at the boundary conditions. The second step is to apply the iterative algorithm of AIM to the equation of motion. We also modify AIM to eliminate the derivatives in each iteration which improves the computational time and precision of the numerical results \cite{Burikham:2017gdm,Cho:2011sf}.

\subsection{Setting the form of differential equation}

By the separation method, the scalar field can be decomposed into
\begin{equation}
	\Phi(t,r,\theta,\varphi) = e^{-i\omega t} \frac{\phi(r)}{r} Y(\theta,\varPhi),
	\label{KG_decompose}
\end{equation} 
where $Y(\theta,\varphi)$ is the spherical harmonic and the gauge field $A_{\mu} = \{-Q/r,0,0,0\}$. The radial scalar field then obeys \cite{Burikham:2017gdm}
\begin{equation}
	 \frac{d^2}{dr_*^2}\phi(r) + (\omega - qQ/r)^2\phi(r) - f(r)\bigg(\frac{l(l+1)}{r^2} + \frac{f'(r)}{r} + m_s^2  \bigg)\phi(r) = 0,
	 \label{KG_equation_Tortoise}
\end{equation}
where $dr_* = dr / f(r)$. The quasinormal frequencies satisfy the boundary conditions (asymptotic behaviours),
\begin{equation}
\phi \sim \begin{cases} 
e^{-i \tilde{\omega} r_*} & \text{as } r \rightarrow r_h \\
e^{i \hat{\omega} r_*} & \text{as } r \rightarrow r_c,
\end{cases}
\label{bdc_phi_}
\end{equation} 
where $\tilde{\omega} = \omega - \frac{qQ}{r_h}$, $\hat{\omega} = \omega - \frac{qQ}{r_c}$, $r_h$ is the event horizon and $r_c$ is the cosmological horizon. If there is no cosmological constant, $r_c \rightarrow \infty$. Performing the radial coordinate transformation $r = 1 / \xi$  \cite{Cho:2011sf,Burikham:2017gdm,Prasia:2016fcc,Moss:2001ga,Ponglertsakul:2018smo}, we obtain
\begin{eqnarray}
	\phi''(\xi) + \frac{p'(\xi)}{p(\xi)}\phi'(\xi) + \frac{(\omega - q Q \xi)^2}{p^2(\xi)}\phi(\xi) 
	- \frac{1}{p(\xi)}\bigg( l(l+1) -\xi \frac{d f(1/\xi)}{d\xi} + \frac{m_s^2}{\xi^2} \bigg) \phi(\xi) = 0,
	\label{KGxi}
\end{eqnarray}
where 
\begin{equation}
	p(\xi) \equiv \xi^2 f(1/\xi).
	\label{p}
\end{equation}
Assume there are $n$ singularities ($\xi_{1}, \xi_{2}, ... \xi_n$) corresponding to $f(1/\xi_i) = 0$. We have to scale out the divergences of the singularities from the radial part of the scalar field. First we apply the boundary condition near the cosmological horizon from Eq.~(\ref{bdc_phi_}) \cite{Moss:2001ga, Prasia:2016fcc, Cho:2009cj}
\begin{equation}
	\phi(\xi) = e^{i \hat{\omega} r_*} u(\xi),   \label{sf0}
\end{equation}
where
\begin{equation}
e^{i \hat{\omega} r_*} \equiv \prod_{i=1}^{n}(\xi - \xi_i)^{\frac{i \hat{\omega}}{2 \kappa_i}},
\end{equation}
and $\kappa_{i}$ is the surface gravity at $\xi_{i}$. The radial part of the scalar field equation Eq.~(\ref{KGxi}) reads
\begin{equation}
	u'' + \frac{(p' - 2 i \hat{\omega})}{p}u'+\frac{\left((\omega - qQ\xi)^2-\hat{\omega}^2\right)}{p^2}u - \frac{1}{p}\Bigg[ l(l+1) -\xi\frac{df(1/\xi)}{d\xi} + \frac{m_s^2}{\xi^2} \Bigg] u = 0.
	\label{KGxi2}
\end{equation}
Next we apply the boundary condition near the event horizon \cite{Moss:2001ga, Prasia:2016fcc, Cho:2009cj} to eliminate the corresponding divergence,
\begin{equation}
	u(\xi) = (\xi - \xi_h)^{\frac{-i \tilde{\omega}}{2\kappa_h}} (\xi - \xi_h)^{\frac{-i \hat{\omega}}{2\kappa_h}} \chi(\xi).  \label{sc2}
\end{equation}
Note the difference in the scaling factors in (\ref{sf0}) and (\ref{sc2}) from Ref.~\cite{Burikham:2017gdm}, the correct $\hat{\omega}~(\tilde{\omega})$ must be used for elimination of the divergence at the cosmological~(event) horizon respectively.  Remarkably, the all-region~(WKB) modes are unchanged and consistent with the values calculated from other methods such as the WKB and analytic formula regardless of the scaling factors at the horizons.  The all-region QNMs are insensitive to the divergences around the horizon for the analysis performed by AIM.  

Finally, Eq.~(\ref{KGxi2}) can be written as \cite{Cho:2011sf,Burikham:2017gdm,Prasia:2016fcc},
\begin{equation}
	\chi''(\xi) = \lambda_0(\xi)\chi'(\xi) + s_0(\xi)\chi(\xi),
	\label{AIMform}
\end{equation}
where $\lambda_0$ and $s_0$ are the coefficients of the equation. With this form, we are ready to use the algorithm of AIM in the next step. 

\subsection{The algorithm of AIM}
From the homogeneous linear second-order differential equation in Eq.~(\ref{AIMform}), AIM uses the asymptotic aspect by taking higher-order differentiation in order to find the a general solution and its eigenvalues.
First, taking differentiation of Eq.~(\ref{AIMform}), we find that \cite{Ciftci:2003,Cho:2011sf}
\begin{equation}
	\chi'''(\xi) = \lambda_1(\xi)\chi'(\xi) + s_1(\xi)\chi(\xi),
\end{equation}
where
\begin{equation}
	\lambda_1(\xi) = \lambda'_0(\xi) + s_0(\xi) + \lambda_0^2(\xi)
\end{equation}
and
\begin{equation}
	s_1(\xi) = s'_0(\xi) + s_0(\xi)\lambda_0(\xi).
\end{equation}
Taking the $n$-time derivation of Eq.~(\ref{AIMform}) to obtain
\begin{equation}
	\chi^{(n+2)}(\xi) = \lambda_n(\xi)\chi'(\xi) + s_n(\xi)\chi(\xi),
\end{equation}
where
\begin{equation}
	\lambda_n(\xi) = \lambda'_{n-1}(\xi) + s_{n-1}(\xi) + \lambda_0(\xi)\lambda_{n-1}(\xi)
	\label{lamdef}
\end{equation}
and
\begin{equation}
	s_n(\xi) = s'_{n-1}(\xi) + s_0(\xi)\lambda_{n-1}(\xi).
	\label{sdef}
\end{equation}
With large enough $n$, the asymptotic of AIM claims that \cite{Cho:2011sf,Burikham:2017gdm,Ciftci:2003}
\begin{equation}
	\frac{s_n(\xi)}{\lambda_n(\xi)} = \frac{s_{n-1}(\xi)}{\lambda_{n-1}(\xi)} \equiv \beta(\xi).
	\label{ASYMP}
\end{equation}
Finally, we get the quantization condition for this method to solve the quasinormal modes,
\begin{equation}
	s_n(\xi) \lambda_{n-1}(\xi) = s_{n-1}(\xi) \lambda_n(\xi).
\end{equation}
One can use this iterative algorithm to find the eigenvalues $\omega$ of Eq.~(\ref{KG_equation_Tortoise}). On the other hand, there is a giant drawback in this algorithm because, in each step, we have to differentiate $s$ and $\lambda$ from the previous step. This causes not just only a lot of time but the divergence of the numerical results. We improve AIM by using Taylor series expansion of $s$ and $\lambda$ around the point $\bar{\xi}$ as follows \cite{Burikham:2017gdm,Cho:2011sf},
\begin{eqnarray}
	\lambda_n(\xi) &=& \sum_{i=0}^{\infty}c_n^i(\xi-\bar{\xi})^i \label{lamb},\\
	s_n(\xi) &=& \sum_{i=0}^{\infty}d_n^i(\xi-\bar{\xi})^i,\label{s}
\end{eqnarray}
where $c_n^i$ and $d_n^i$ are the $i^{th}$ Taylor coefficient's of $\lambda_n(\xi)$ and $s_n(\xi)$ respectively. Substituting these Taylor expansions, Eq.~(\ref{lamb}) and Eq.~(\ref{s}), into Eq.~(\ref{lamdef}) and Eq.~(\ref{sdef}) leads to
\begin{eqnarray}
	c_n^i &=& (i+1)c_{n-1}^{i+1} + d_{n-1}^i + \sum_{k=0}^{i}c_0^kc_{n-1}^{i-k}, \label{cfinal} \\
	d_n^i &=& (i+1)d_{n-1}^{i+1} + \sum_{k=0}^{i}d_0^kc_{n-1}^{i-k}.
\end{eqnarray}
The quantization condition can be written in terms of these coefficients,
\begin{equation}
	d_n^0 c_{n-1}^0 - d_{n-1}^0 c_n^0 = 0.
\end{equation}
With this improvement of AIM, there is no derivative operator in the iteration process. We can apply the recursive method to find the quasinormal modes. 

\section{Numerical results}\label{secNUM}

In this section, we numerically calculate the quasinormal frequencies of the charged scalar in the generalized black hole background by AIM. Generically, there are three kinds of QNMs that can be found by AIM; near event horizon, near cosmological horizon and all-region~(WKB) modes \cite{Burikham:2017gdm}. We consider the near horizon QNMs in 4 cases; the QNMs of the near extremal Schwarzschild de-Sitter dRGT black hole in subsection \ref{subNEEXSCHW}, the QNMs of the near extremal Reissner-Nordstr\"{o}m de-Sitter dRGT black hole in subsection \ref{subNEEXRN}, the QNMs of the non-extremal Schwarzschild de-Sitter dRGT black hole in subsection \ref{subNONEXSCHW}, and the QNMs of the non-extremal Reissner-Nordstr\"{o}m de-Sitter dRGT black hole in subsection \ref{subNONEXRN}.

The results of QNMs in this section are calculated at two $\bar{\xi}$ expansion points. The first point is near the event horizon, $\bar{\xi}_{1} = 0.999 \xi_h = 0.999 / r_h$. The second point is near the cosmological horizon, $\bar{\xi}_{2} = 1.001 \xi_c = 1.001 / r_c$.

\subsection{The Near-Extremal Schwarzschild de-Sitter dRGT cases}\label{subNEEXSCHW}

In this case, the equation of motion of the scalar field is (Eq.~(\ref{KG_equation_Tortoise}) with zero charge) \cite{Burikham:2017gdm}
\begin{equation}
\frac{d^2}{dr_*^2}\phi(r) + \omega ^2 \phi(r) - f(r)\bigg(\frac{l(l+1)}{r^2} + \frac{f'(r)}{r} + m_s^2  \bigg)\phi(r) = 0,
\label{KG_equation_SCHW}
\end{equation}
which
\begin{equation}
f(r) = 1 - \frac{2M}{r} - \frac{\Lambda r^2}{3} + \gamma r + \zeta,
\label{fr_SCHW}
\end{equation}
where $M$ is the mass of the black hole, $\Lambda$ is the cosmological constant, $\gamma$ and $\zeta$ are the coefficients from the dRGT effects.
The condition for the near extremal black hole is
\begin{equation}
\frac{r_c - r_h}{r_h} \ll 1.
\label{cond_extrm}
\end{equation}
To consider QNMs of the near-extremal black hole, we set parameters, corresponding to the condition Eq.~(\ref{cond_extrm}), as $M = 1, \Lambda = 5.0001, \zeta = 2.51465, l = 0, \gamma = 0.05, \text{and } m_s = 0.00$. Then the event horizon is at $r_{h}=0.8460$ and the cosmological horizon is at $r_{c}=0.8509$. 

\begin{table}[t]
	\centering
	\begin{tabular}{|c|c|c|}
		\hline
		~~$n$~~& QNMs($\omega_n$) by AIM & $\Delta \omega_n $ by AIM \\
		\hline
		0 & ~~1.41 $\times 10^{-7}$ i~~ &  -   \\
		\hline
		1 & -0.01216 i & ~~ -0.01216 i ~~ \\
		\hline
		2 & -0.02433 i & ~~ -0.01216 i ~~ \\
		\hline
	\end{tabular}
	\caption{The quasinormal modes of uncharged near-extremal dRGT black hole with 100 iterations of AIM for the parameters, $M = 1, \Lambda = 5.0001, \zeta = 2.51465, l = 0, \gamma = 0.05, \text{and } m_s = 0.00$ at $\bar{\xi}_{1}$. $\kappa_h = 0.01216 \sim -\kappa_c$.}
	\label{ExtSCHW1}
\end{table}

As shown in Table \ref{ExtSCHW1}, the gaps between the overtones of the near-horizon QNMs are given precisely by $\Delta \omega = i\kappa_h$.

\subsection{The Near-Extremal Reissner-Nordstr\"om de-Sitter dRGT cases}\label{subNEEXRN}

The equation of motion for the zero-charge scalar field in the charged black hole spacetime is Eq.~(\ref{KG_equation_SCHW}) with the metric~\cite{Burikham:2017gdm}
\begin{equation}
f(r) = 1 - \frac{2M}{r} + \frac{Q^2}{r^2}- \frac{\Lambda r^2}{3} + \gamma r + \zeta,
\label{fr_RN}
\end{equation}
where $Q$ is the black hole charge.

We set the spacetime parameters to obtain two near-extremal cases; the event horizon is very close to the cosmological horizon~(small universe), and the event horizon is very close to the Cauchy horizon. The resulting QNMs are shown in Table \ref{ExtdS1} and Table \ref{ExtdS2} respectively.
\begin{table}[h]
	\centering
	\begin{tabular}{|c|c|c|}
		\hline
		~~$n$~~& QNMs($\omega_n$) by AIM & Im($\Delta \omega_n $) by AIM\\
		\hline
		0 & ~~4.27 $\times 10^{-15}$ i~~ & ~~ - ~~ \\
		\hline
		1 & -0.0006779 i & ~~ -0.0006779 i ~~ \\
		\hline
		2 & -0.001356 i & ~~ -0.0006779 i ~~  \\
		\hline
	\end{tabular}
	\caption{The quasinormal modes of charged near-extremal dRGT black hole with 100 iterations of AIM for the parameters, $M = 1, Q = 0.5, \Lambda = 0.09, q = 0.00, \zeta = 0.00, l = 0, \gamma = -0.030408, \text{and } m_s = 0.01$ at the point $\bar{\xi}_{1}, \kappa_h = 0.0006779$.}
	\label{ExtdS1}
\end{table}

Table \ref{ExtdS1} shows the near-horizon QNMs by AIM for the 3 lowest frequencies. The parameters are chosen such that the extremal condition Eq. (28) is satisfied and the point we solve using AIM is near the event horizon $\bar{\xi}_{1}=0.999\xi_{h}$. The numerical results of the QNMs are still a tower pattern where the gap in the imaginary parts is equal to the surface gravity at the horizon.

\begin{table}[h]
	\centering
	\begin{tabular}{|c|c|c|}
		\hline
		~~$n$~~ &~~ QNMs($\omega_n$) by AIM ~~& ~~ Im($\Delta \omega_n $) by AIM ~~ \\
		\hline
		0 & ~~ 3.022 $\times 10^{-7}$ i ~~& ~~ - ~~ \\
		\hline
		1 & -0.009264 i & -0.009264 i \\
		\hline
		2 & -0.01852 i & -0.009264 i \\
		\hline
	\end{tabular}
	\caption{The quasinormal modes of charged near-extremal dRGT black hole with 200 iterations of AIM for the parameters, $M = 1, Q = 0.8805, \Lambda = 0.03, q = 0.00, \zeta = 0.00, l = 2, \gamma = 0.4, \text{and } m_s = 0.01$ at the point $\bar{\xi}_{1}, \kappa = 0.009264$.}
	\label{ExtdS2}
\end{table}

We consider another case of the near extremal black hole; when the event horizon is close to the Cauchy horizon or $r_- \sim r_h$. The quasinormal frequencies generated by AIM are presented in Table \ref{ExtdS2}. Again, the gaps between frequencies of the near-horizon QNMs are precisely equal to the value of the surface gravity at the horizon.  This is similar to the behaviour of the all-region WKB modes of the near-extremal black hole~(see also e.g. Ref.~\cite{Ponglertsakul:2018smo} for the black string version with almost identical formula and derivation to the black hole case).

\subsection{The Non-Extremal Schwarzschild de-Sitter dRGT cases} \label{subNONEXSCHW}

In this section, we numerically calculate the near-horizon QNMs of the scalar in the non-extremal Schwarzschild de-Sitter background for both event and cosmological horizon modes. Again, we choose to observe at the points $\bar{\xi}_{1} = 0.999\xi_h = 0.999/r_h$ (near the event horizon) and $\bar{\xi}_{2} = 1.001\xi_c = 1.001 / r_c$ (near the cosmological horizon). 

The dynamic of the scalar field follows the Eq.~(\ref{KG_equation_SCHW}) with the same function $f(r)$ in Eq.~(\ref{fr_SCHW}). However, the set of parameter does not satisfy the condition in Eq.~(\ref{cond_extrm}). To investigate the pattern of the tower in QNMs near the horizons, we vary all physical parameters in Eq.~(\ref{KG_equation_SCHW}).

\begin{table}[h]
	\centering
	\begin{tabular}{|c|c|c|c|c|c|c|c|}
		\hline
		$\Lambda$ &~~$n$~~ &~~ $\omega (\text{at } \bar{\xi}_{1})$ ~~& ~~ $\omega (\text{at } \bar{\xi}_{2})$ ~~ & ~~ Im($\Delta \omega_n (\text{at } \bar{\xi}_{1}))$ ~~ & ~~ Im($\Delta \omega_n (\text{at } \bar{\xi}_{2}))$ ~~ & $\kappa_h$ & $\kappa_c$\\
		\hline
		~0.0005~ & 0 &~~8.42$\times 10^{-8}$ i~~&~~6.73$\times 10^{-11}$ i ~~& -& -& &\\
		& 1 &-1.265 i & -0.4002 i & -1.265 i & -0.4002 i &~~1.265~~&~-0.4002 ~\\
		& 2 &-2.529 i & -0.8004 i & -1.265 i & -0.4002 i & &\\
		& 3 &-3.794 i & -1.2006 i & -1.265 i & -0.4002 i & &\\
		\hline					
		~0.005~ & 0 &3.74$\times 10^{-8}$ i & -2.13$\times 10^{-12}$ i & -& -& &\\
		& 1 &-1.262 i & -0.4021 i & -1.262 i & -0.4021 i & 1.262 & -0.4021 \\
		& 2 &-2.524 i & -0.8041 i & -1.262 i & -0.4021 i & &\\
		& 3 &-3.786 i & -1.2062 i & -1.262 i & -0.4021 i & &\\
		\hline
		~0.05~ & 0 &-5.56$\times 10^{-8}$ i & -6.20$\times 10^{-11}$ i & -& -& &\\
		& 1 &-1.235 i & -0.4191 i & -1.235 i & -0.4191 i & 1.235 & -0.4191 \\
		& 2 &-2.469 i & -0.8382 i & -1.235 i & -0.4191 i & &\\
		& 3 &-3.704 i & -1.2573 i & -1.235 i & -0.4191 i & &\\
		\hline
		~0.5~ & 0 &-6.76$\times 10^{-8}$ i & -3.41$\times 10^{-11}$ i & -& -& &\\
		& 1 &-0.9343 i & -0.4826 i & -0.9343 i & -0.4826 i & 0.9343 & -0.4826 \\
		& 2 &-1.8687 i & -0.9653 i & -0.9343 i & -0.4826 i & &\\
		& 3 &-2.8030 i & -1.4479 i & -0.9343 i & -0.4826 i & &\\
		\hline
	\end{tabular}
	\caption{The quasinormal modes of non-extremal uncharged dRGT black hole with 100 iterations of AIM for the parameters, $M = 1, \gamma = 0.8, \zeta = 0, l = 0, \text{and } m_s = 0.00$ around the point $\bar{\xi}_{1}$ and $\bar{\xi}_{2}$.}
	\label{nonexvarylambdaCHW}
\end{table}

Variation of cosmological constant $\Lambda$ has effect on the gaps between the QNMs as shown in Table \ref{nonexvarylambdaCHW}.  The patterns of tower remain at the points near both the event and cosmological horizons.

\begin{table}[h]
	\centering
	\begin{tabular}{|c|c|c|c|c|c|c|c|}
		\hline
		$\gamma$ &~~$n$~~ &~~ $\omega (\text{at } \bar{\xi}_{1})$ ~~& ~~ $\omega (\text{at } \bar{\xi}_{2})$ ~~ & ~~ Im($\Delta \omega_n (\text{at } \bar{\xi}_{1}))$ ~~ & ~~ Im($\Delta \omega_n (\text{at } \bar{\xi}_{2}))$ ~~ & $\kappa_h$ & $\kappa_c$\\
		\hline
		~-0.1~ & 0 &~~-7.88$\times 10^{-11}$ i~~&~~-1.47$\times 10^{-10}$ i ~~& -& -& &\\
		& 1 &-0.07847 i & -0.03173 i & -0.07847 i & -0.03173 i &~~0.07847~~&~-0.03173 ~\\
		& 2 &-0.1569 i & -0.06345 i & -0.07847 i & -0.03173 i & &\\
		& 3 &-0.2354 i & -0.09518 i & -0.07847 i & -0.03173 i & &\\
		\hline					
		~0.0~ & 0 &-5.90$\times 10^{-10}$ i & -1.20$\times 10^{-9}$ i & -& -& &\\
		& 1 &-0.2487 i & -0.01757 i & -0.2487 i & -0.01757 i & 0.2487 & -0.01757 \\
		& 2 &-0.4973 i & -0.03514 i & -0.2487 i & -0.01757 i & &\\
		& 3 &-0.7460 i & -0.05270 i & -0.2487 i & -0.01757 i & &\\
		\hline
		~0.1~ & 0 &-3.61$\times 10^{-10}$ i & 4.61$\times 10^{-9}$ i & -& -& &\\
		& 1 &-0.3916 i & -0.0532 i & -0.3916 i & -0.0532 i & 0.3916 & -0.0532 \\
		& 2 &-0.7832 i & -0.1064 i & -0.3916 i & -0.0532 i & &\\
		& 3 &-1.1749 i & -0.1596 i & -0.3916 i & -0.0532 i & &\\
		\hline
	\end{tabular}
	\caption{The quasinormal modes of uncharged non-extremal dRGT black hole with 100 iterations of AIM for the parameters, $M = 1, \Lambda = 0.001, \zeta = 0, l = 2, \text{and } m_s = 0.2$ around the point $\bar{\xi}_{1}$ and $\bar{\xi}_{2}$.}
	\label{nonexvarygammaCHW}
\end{table}

Next, the numerical results from the variation of the coefficient in the linear term $\gamma$ at point $\bar{\xi}_{1}$ and $\bar{\xi}_{2}$ are listed in Table \ref{nonexvarygammaCHW}. In this case, we set the mass of the scalar field and the angular momentum to be nonzero~($m_s = 0.2, l= 2$). The numerical values of QNMs in these tables have vanishing real parts. The imaginary parts have constant gaps which can be well approximated by the values of the corresponding surface gravity. 

The scalar mass has no effect on the function $f(r)$, thus the surface gravity near the event and cosmological horizons $\kappa_{h}, \kappa_{c}$ do not change under the the scalar mass variation. From the results in the Table \ref{nonexvarymassCHW}, there is again a tower pattern in the QNMs near the event and cosmological horizon. 

\begin{table}[h]
	\centering
	\begin{tabular}{|c|c|c|c|c|c|c|c|}
		\hline
		$m_s$ &~~$n$~~ &~~ $\omega (\text{at } \bar{\xi}_{1})$ ~~& ~~ $\omega (\text{at } \bar{\xi}_{2})$ ~~ & ~~ Im($\Delta \omega_n (\text{at } \bar{\xi}_{1}))$ ~~ & ~~ Im($\Delta \omega_n (\text{at } \bar{\xi}_{2}))$ ~~ & $\kappa_h$ & $\kappa_c$\\
		\hline
		~0.001~ & 0 &~~2.44$\times 10^{-11}$ i~~&~~-6.32$\times 10^{-12}$ i ~~& -& -& &\\
		& 1 &-0.5238 i & -0.1048 i & -0.5238 i & -0.1048 i &~~0.5238~~&~-0.1048 ~\\
		& 2 &-1.0476 i & -0.2096 i & -0.5238 i & -0.1048 i & &\\
		& 3 &-1.5713 i & -0.3144 i & -0.5238 i & -0.1048 i & &\\
		\hline					
		~0.01~ & 0 &~~2.44$\times 10^{-11}$ i~~&~~-6.32$\times 10^{-12}$ i ~~& -& -& &\\
		& 1 &-0.5238 i & -0.1048 i & -0.5238 i & -0.1048 i &~~0.5238~~&~-0.1048 ~\\
		& 2 &-1.0476 i & -0.2096 i & -0.5238 i & -0.1048 i & &\\
		& 3 &-1.5713 i & -0.3144 i & -0.5238 i & -0.1048 i & &\\
		\hline
		~0.1~ & 0 &~~2.44$\times 10^{-11}$ i~~&~~-6.32$\times 10^{-12}$ i ~~& -& -& &\\
		& 1 &-0.5238 i & -0.1048 i & -0.5238 i & -0.1048 i &~~0.5238~~&~-0.1048 ~\\
		& 2 &-1.0476 i & -0.2096 i & -0.5238 i & -0.1048 i & &\\
		& 3 &-1.5713 i & -0.3144 i & -0.5238 i & -0.1048 i & &\\
		\hline
	\end{tabular}
	\caption{The quasinormal modes of uncharged non-extremal dRGT black hole with 100 iterations of AIM for the parameters, $M = 1, \Lambda = 0.003, \zeta = 0, l = 1, \text{and } \gamma = 0.2$ around the point $\bar{\xi}_{1}$ and $\bar{\xi}_{2}$.}
	\label{nonexvarymassCHW}
\end{table}

\subsection{The Non-Extremal Reissner-Nordstr$\ddot{{\rm o}}$m de-Sitter dRGT cases} \label{subNONEXRN}

\begin{table}[t]
	\centering
	\noindent\makebox[\textwidth]{%
	\begin{tabular}{|c|c|c|c|c|c|c|c|}
		\hline
		$\Lambda$ &~~$n$~~ &~~ $\omega (\text{at } \bar{\xi}_{1})$ ~~& ~~ $\omega (\text{at } \bar{\xi}_{2})$ ~~ & ~~ $qQ/r_h$ ~~ & ~~ $qQ/r_c$ ~~ & $\kappa_h$ & $\kappa_c$\\
		\hline
		~0.001~ & 0 &~~0.0541 - 1.3$\times10^{-7}$ i~~&~~5.90$\times 10^{-4}$ - 3.0$\times 10^{-11}$ i ~~& & & &\\
		& 1 &0.0541 - 0.3216 i & 5.90$\times 10^{-4}$ - 0.0308 i & ~0.0541~ & ~5.90$\times 10^{-4}$~ &~~0.3216~~&~- 0.0308 ~\\
		& 2 &0.0541 - 0.6432 i & 5.90$\times 10^{-4}$ - 0.0617 i &  &  & &\\
		& 3 &0.0541 - 0.9648 i & 5.90$\times 10^{-4}$ - 0.0926 i &  &  & &\\
		\hline					
		~0.01~ & 0 &~0.05365 + 5.6$\times 10^{-8}$ i~ & 0.00385 + 1.2$\times 10^{-10}$ i & & & &\\
		& 1 &0.05365 - 0.3109 i & 0.00385 - 0.0593 i & 0.05365 & 0.00385 & 0.3109 & -0.0593 \\
		& 2 &0.05365 - 0.6219 i & 0.00385 - 0.1186 i &  &  & &\\
		& 3 &0.05365 - 0.9328 i & 0.00385 - 0.1779 i &  &  & &\\
		\hline
		~0.1~ & 0 &0.04763 + 2.6$\times 10^{-9}$ i & 0.01948 + 1.1$\times 10^{-8}$ i & & & &\\
		& 1 &0.04763 - 0.1861 i & 0.01948 - 0.1058 i & 0.04763 & 0.01948 & 0.1861 & -0.1058 \\
		& 2 &0.04763 - 0.3722 i & 0.01948 - 0.2115 i &  &  & &\\
		& 3 &0.04763 - 0.5582 i & 0.01948 - 0.3173 i &  &  & &\\
		\hline
	\end{tabular}}
	\caption{The quasinormal modes of charged non-extremal dRGT black hole with 100 iterations of AIM for the parameters, $M = 1, Q = 0.1, q = 0.99, \zeta = 0, l = 0, m_s = 0.0 \text{ and } \gamma = 0.05$ around the point $\bar{\xi}_{1}$ and $\bar{\xi}_{2}$.}
	\label{nonexvaryLambdaRN}
\end{table}

\begin{table}[t]
	\centering
	\begin{tabular}{|c|c|c|c|c|c|c|c|}
		\hline
		$\gamma$ &~~$n$~~ &~~ $\omega (\text{at } \bar{\xi}_{1})$ ~~& ~~ $\omega (\text{at } \bar{\xi}_{2})$ ~~ & ~~ $qQ/r_h$ ~~ & ~~ $qQ/r_c$ ~~ & $\kappa_h$ & $\kappa_c$\\
		\hline
		~-0.05~ & 0 &~~0.04302 + 6.8$\times10^{-8}$ i~~&~~0.01026 - 7.4$\times 10^{-7}$ i ~~& & & &\\
		& 1 &0.04302 - 0.1554 i & 0.01026 - 0.04623 i & ~0.04302~ & 0.01026 &~~0.1554~~&~-0.04643 ~\\
		& 2 &0.04302 - 0.3108 i & 0.01026 - 0.09290 i &  &  & &\\
		& 3 &0.04302 - 0.4661 i & 0.01026 - 0.1393 i &  &  & &\\
		\hline					
		~0.00~ & 0 &0.04894 + 8.5$\times 10^{-8}$ i & 0.006104 - 1.8$\times 10^{-7}$ i & & & &\\
		& 1 &0.04894 - 0.2364 i & 0.006104 - 0.05020 i & 0.04894 & 0.006104 & 0.2364 & -0.05025 \\
		& 2 &0.04894 - 0.4730 i & 0.006104 - 0.1005 i &  &  & &\\
		& 3 &0.04894 - 0.7095 i & 0.006104 - 0.1508 i &  &  & &\\
		\hline
		~0.05~ & 0 &0.05365 - 5.6$\times 10^{-8}$ i & 0.003845 + 1.3$\times 10^{-7}$ i & & & &\\
		& 1 &0.05365 - 0.3109 i & 0.003845 - 0.05930 i & 0.05365 & 0.003845 & 0.3109 & -0.05931 \\
		& 2 &0.05365 - 0.6219 i & 0.003845 - 0.1186 i &  &  & &\\
		& 3 &0.05365 - 0.9328 i & 0.003845 - 0.1779 i &  &  & &\\
		\hline
	\end{tabular}
	\caption{The quasinormal modes of charged non-extremal dRGT black hole with 100 iterations of AIM for the parameters, $M = 1, Q = 0.1, \Lambda = 0.01, q = 0.99, \zeta = 0, l = 2 \text{ and } m_s = 1.00$ around the point $\bar{\xi}_{1}$ and $\bar{\xi}_{2}$.}
	\label{nonexvaryGammaRN}
\end{table}

\begin{table}[t]
	\centering
	\begin{tabular}{|c|c|c|c|c|c|c|c|}
		\hline
		$Q$ &~~$n$~~ &~~ $\omega (\text{at } \bar{\xi}_{1})$ ~~& ~~ $\omega (\text{at } \bar{\xi}_{2})$ ~~ & ~~ $qQ/r_h$ ~~ & ~~ $qQ/r_c$ ~~ & $\kappa_h$ & $\kappa_c$\\
		\hline
		~-0.2~ & 0 &~~-0.05462 - 1.6 $\times 10^{-8}$ i~~&~~-0.00388 + 1.1 $\times 10^{-9}$ i ~~& & & &\\
		& 1 &-0.05462 - 0.3107 i & -0.00388 - 0.0593 i & ~-0.05462~ & -0.00388 &~~0.3107~~&~-0.0593 ~\\
		& 2 &-0.05462 - 0.6213 i & -0.00388 - 0.1186 i &  &  & &\\
		& 3 &-0.05462 - 0.9320 i & -0.00388 - 0.1779 i &  &  & &\\
		\hline
		~0.5~ & 0 &0.1451 + 2.4$\times 10^{-7}$ i & 0.009708 + 1.4$\times 10^{-8}$ i & & & &\\
		& 1 &0.1451 - 0.3071 i & 0.009708 - 0.0593 i & 0.1451 & 0.009708 & 0.3071 & -0.0593 \\
		& 2 &0.1451 - 0.6142 i & 0.009708 - 0.1187 i &  &  & &\\
		& 3 &0.1451 - 0.9212 i & 0.009708 - 0.1780 i &  &  & &\\
		\hline
	\end{tabular}
	\caption{The quasinormal modes of uncharged non-extremal dRGT black hole with 100 iterations of AIM for the parameters, $M = 1, \Lambda = 0.01, \gamma = 0.05, q = 0.5, \zeta = 0.0, l = 1 \text{ and } m_s = 0.50$ around the point $\bar{\xi}_{1}$ and $\bar{\xi}_{2}$.}
	\label{nonexvaryChargeRN}
\end{table}

The equation of motion for the charged scalar field in the non-extremal Reissner-Nordstr$\ddot{{\rm o}}$m de-Sitter dRGT black hole satisfies Eq.~(\ref{KG_equation_Tortoise}) with $f(r)$ given by Eq.~(\ref{fr_RN}).

The structures of tower have been found in the near-horizon QNMs for this case in Ref.~\cite{Burikham:2017gdm}~(with errors in the real parts by a factor of $1/2$). The real parts of QNMs are shifted by the Coulomb interaction between the charged scalar and the charged black hole by an amount of $qQ/r_{h}$ and $qQ/r_{c}$ for the event and cosomological horizon respectively. We vary parameters such as cosmological constant $\Lambda$, the linear term $\gamma$, the charge of black hole $Q$, and the mass of scalar field $m_s$.  The results are presented in Table \ref{nonexvaryLambdaRN}-\ref{nonexvaryChargeRN}. In each case, we report the numerical results of the first four quasinormal frequencies, the difference between overtones in the imaginary parts, and the values of the surface gravities. We choose $\bar{\xi}_{1} = 0.999 \xi_h$, and $\bar{\xi}_{2} = 1.001 \xi_c$ to be points on spacetime for our observation of the quasinormal frequencies.

In Table \ref{nonexvaryLambdaRN}, we present the QNMs near the event and cosmological horizons by the variation of the cosmological constant from $0.001$ to $0.1$. The magnitudes of the real part equal $qQ/r_i$ where $r_i$ is either the event horizon or the cosmological horizon. Moreover, the gaps between overtones of QNMs converge to the surface gravities.

In Table \ref{nonexvaryLambdaRN}, we consider the quasinormal frequencies at the points $\bar{\xi}_{1} = 0.999 \xi_h$ and $\bar{\xi}_{2} = 1.001 \xi_c$ and vary the coefficient of the linear term from the effect of dRGT. The structures of tower appear in the numerical result of QNMs where the real parts shift with the value $qQ/r_i$ and the gaps between overtones are of the surface gravities.

The sign of the real parts of the quasinormal frequencies depends on the sign of the black hole charge. As shown in Table \ref{nonexvaryChargeRN}, the real parts of the QNMs are indeed given by $qQ/r_i$.  Furthermore, the gaps of the imaginary parts are again equal to the surface gravities. The structure of tower is still found in the numerical results.

With variations of the 4 parameters for general cases displayed in Table \ref{nonexvaryLambdaRN}-\ref{nonexvaryChargeRN}, we do not only get the structures of tower for QNMs but also learn that the gaps in imaginary parts of the overtones are equal the corresponding suface gravities.  This is not surprising since perturbations of fields near the horizons should be governed entirely by the physical parameters of the horizons and fields~($q$ but not $m_{s}, l$). 

\section{The exact solutions of near-horizon quasinormal frequencies}\label{sec_exact}

According to the notable numerical results of the QNMs of the black holes near the event and cosmological horizons in the previous section, it is expected that the quasinormal frequencies are generically forming the tower patterns in the vicinity of the horizons. In this section, we analytically calculate the exact solutions of quasinormal frequencies near the event and cosmological horizons in general static spherically symmetric spacetime.  This is an extension of the analysis in Ref.~\cite{Li:2017gwm} to include the effect of gauge field on the charged scalar.  The results are exact and universal, the Coulomb potential shifts the real parts of the QNMs by $qQ/r_{a}$ for $r_{a}=(r_{h}, r_{c})$ of the event and cosmological horizon respectively.

\subsection{Exact solution of the QNMs near the event and cosmological horizons in the static spherically symmetric spacetime}\label{subsection_general_coordinate}

The evolution of charged scalar field on the spherically symmetric spacetime can be described by Eq. (\ref{KG_equation_Tortoise}). It is convenient to redefine the radial function such that $\phi(r)=rR(r)$. Then Eq. (\ref{KG_equation_Tortoise}) takes the form
\begin{align}
\frac{d^2 R(r)}{dr^2} + \bigg( \frac{f'(r)}{f(r)} + \frac{2}{r} \bigg)\frac{d R(r)}{dr} + \bigg[ \frac{1}{f^2(r)}\bigg(\omega - \frac{qQ}{r}\bigg)^2 - \frac{l(l+1)}{f(r)r^2} - \frac{m_s^2}{f(r)} \bigg]R(r)=0,
\label{KGequationR}
\end{align}
where in this case $f(r)$ is a generic function of $r$ with at least two singular points i.e. $f(r_h)=f(r_c)=0$. We now define a new coordinate 
\begin{equation}
z = 1 - \frac{r_a}{r},
\label{defZ}
\end{equation}
where $a=\{h,c\}$ associates with each horizon. Therefore in this coordinate system, the horizons are located at $z=0$. Near each horizon, the metric function can be approximated as $f\approx f'_0 z$ where $f'_0$ is $\displaystyle{\frac{df}{dz}\Big{|}_{z=0}}$. Hence the scalar field equation (\ref{KGequationR}) can be rewritten in the near horizon limit as
\begin{align}
\frac{d^2R(z)}{dz^2} + \frac{1}{z}\frac{dR(z)}{dz} + \frac{r_a^2}{\left(1-z\right)^4}\left[\frac{\left(\omega-\frac{qQ(1-z)}{r_a}\right)^2}{f'^2_0 z^2} - \frac{m_s^2}{f'_0 z} - \frac{l(l+1)\left(1-z\right)^2}{f'_0 z r_a^2}\right]R(z) &= 0.\label{KGequationZ}
\end{align}
We can simplify this equation by introducing the following radial field function
\begin{align}
R(z) &= z^{\alpha}(1-z)^{\beta}F(z).
\end{align}
By substituting this into Eq. (\ref{KGequationZ}), we find that $F(z)$ satisfies the standard hypergeometric differential equation
\begin{align}
z(1-z)\frac{d^2F(z)}{dz^2} + \left[c-\left(a+b+1\right)z\right]\frac{dF(z)}{dz} - ab F(z) &= 0,  \label{hypergeomtric_diff}
\end{align}
where we have taken $z$ to be small and ignored the higher order power of $z$. The power factor $\alpha$ and $\beta$ are constrained to be 
\begin{equation}
\alpha = \pm \frac{\left(\omega r_a - q Q\right)i}{2 r_a \kappa_a},
\end{equation}
\begin{align}
\beta &= \frac{1}{2}\pm\frac{1}{2\kappa_{a}r_{a}}\sqrt{\Big[r_{a} \Big(\kappa_{a} \left(2 l (l+1)+r_{a} \left(\kappa_{a}+6 \mu^2 r_{a}\right)\right)-6 r_{a} \omega ^2\Big)-q^2 Q^2+6 q Q r_{a} \omega \Big]},
\end{align}
where the surface gravity is defined as $\kappa_a = f'_{0}/2 r_a$. The hypergeometric parameters can be defined as the following
\begin{align}
a &= \alpha+\beta - \delta, \label{eqa}\\
b &= \alpha+\beta + \delta, \label{eqb}\\
c &= 1+2\alpha, \label{eqc}
\end{align}
where 
\begin{align}
\delta &= \frac{\sqrt{r_{a} \left(2 q Q \omega +4 \kappa_{a} m_{s} ^2 r_{a}^2-3 r_{a} \omega ^2\right)}}{2\kappa_{a} r_{a}}.
\end{align}
Thus for $z\to 0$, the general solution of Eq.(\ref{KGequationZ}) has the form
\begin{align}
R(z) &= A_1 z^{\alpha} (1-z)^{\beta} F\left(a,b,c,z\right) + A_2 z^{1-c+\alpha}(1-z)^{\beta}F\left(1+a-c,1+b-c,2-c,z\right), \label{gensol}
\end{align}
where $F(a,b,c,z)$ is the hypergeometric function and $A_1$ and $A_2$ are arbitrary constant for non-integer $c$. 

However if $c$ is an integer $n$, this gives an exact formula for the discrete frequencies as
\begin{align}
\omega &= \frac{qQ}{r_a} \pm i\kappa_a\left(n-1\right),
\end{align}
the plus sign corresponds to choosing negative value of $\alpha$.  The solutions have different forms for positive and negative integer cases.  If $c$ is a positive integer, the general solution around $z=0$ becomes
\begin{align}
R(z) &= z^{\alpha}(1-z)^{\beta}\left[A_1 F(a,b,c,z) + A_2 H(a,b,c,z)\right]. \label{solp}
\end{align}
For $c$ is zero and negative integer, the general solution is given by
\begin{align}
R(z) &= z^{-\alpha}(1-z)^{\beta}\left[A_1 F(1+a-c,1+b-c,2-c,z) + A_2 H(1+a-c,1+b-c,2-c,z)\right]. \label{soln}
\end{align}
The second linearly independent solution $H$ is defined such that
\begin{align}
H(a,b,c,z) &= F(a,b,c,z)\ln z - \sum\limits_{k=1}^{c-1}\frac{(c-1)!(k-1)!}{(c-k-1)!(1-a)_k(1-b)_k}\left(-z\right)^{-k} \\
&~~~+ \sum\limits_{k=0}^{\infty}\frac{(a)_k(b)_k}{(c)_k k!}z^k \left[\psi(a+k)+\psi(b+k)-\psi(1+k)-\psi(c+k)\right],
\end{align}
where $\psi(z)$ is the digamma function.  In this case, the QNMs for minus sign choice~(plus sign for cosmic horizon) of $\alpha$ are given by
\begin{align}
\omega &= \frac{qQ}{r_a} + i|\kappa_a|\left(n-1\right),\quad\quad n=0,-1,-2,...  \label{anao}
\end{align}
This is exactly the modes found numerically by AIM.  The mode with zero imaginary part is also obtained when $c=1$ in the positive integer case.    

\subsection{Exact solution of the QNMs in all-region modes of the linearly approximated metric}\label{subsection_general_coordinate_far_from_horizon}

In general, the quasinormal frequencies can be categorized into three kinds which are the near event horizon, the near cosmological horizon, and the all-region modes~\cite{Burikham:2017gdm}. In this section, we derive analytic formulae in terms of the roots of quartic equation~(though they will not be presented due to the lengthy expressions) for the quasinormal frequencies of the all-region modes in the near-horizon coordinates. These are the modes that are derived in the near-horizon approximation and yet can reach the asymptotically far region. Since the metric is linearly approximated, the modes naturally are not equal to the all-region QNMs found by the WKB method or AIM in the usual coordinates without approximation.

As mentioned earlier in section \ref{subsection_general_coordinate}, Eq.~(\ref{hypergeomtric_diff}) can be regarded as a hypergeometric differential equation in the $z\to 0$ limit. The general solution of this equation can be expressed as
\begin{align}
R(z) &= A_1 z^{\alpha} (1-z)^{\beta} F\left(a,b,c,z\right) + A_2 z^{1-c+\alpha}(1-z)^{\beta}F\left(1+a-c,1+b-c,2-c,z\right),
\end{align}
where $A_{1} \text{ and } A_{2}$ are arbitrary constants. $a$, $b$, $c$ are defined as in Eq.~(\ref{eqa}--\ref{eqc}),
and we choose 
\begin{eqnarray}
\alpha &=& \frac{i (q Q-r_a \omega )}{2\kappa_{a} r_a} \nonumber \\
\beta &=& \frac{1}{2} + \frac{\sqrt{2 \kappa _a r_a \left(3 r_a^2 m_s^2+l^2+l\right)+6 \omega  r_a \left(q Q-\omega  r_a\right)+ \kappa _a^2 r_a^2- q^2 Q^2}}{2 \kappa _a r_a}.
\end{eqnarray}
Thus near the event horizon, we must choose $A_{2}=0$. Then the solution satisfying the boundary condition at the horizon becomes
\begin{equation}
	R(z) = A_1 z^{\alpha} (1-z)^{\beta} F\left(a,b,c,z\right).
\end{equation}
In order to satisfy the outgoing-wave boundary condition at large $r$ i.e. $z\to 1$, we make use of the following transformation~\cite{abramowitz+stegun},
\begin{eqnarray}
	F(a,b,c,z) =&&\frac{\Gamma(c)\Gamma(c-a-b)}{\Gamma(c-a)\Gamma(c-b)}F(a,b,a+b-c+1,1-z) \nonumber\\
	&&+ (1-z)^{c-a-b} \frac{\Gamma(c)\Gamma(a+b-c)}{\Gamma(a)\Gamma(b)}F(c-a,c-b,c-a-b+1,1-z).
\end{eqnarray}
At large $r$, only outgoing wave is allowed. Thus we must have the second term on the RHS vanishes
\begin{align}
a &= -n, \nonumber \\
\text{or}~~~~b &= -n,~~~~~~~~~~~~~~~n=0,1,2,... \label{allw}
\end{align}
Eq.~(\ref{allw}) is a quartic equation of $\omega$ and there are four roots of the QNMs.  These four roots form four towers of the QNMs as we will numerically demonstrate subsequently.

\begin{figure}[h]
	\includegraphics[width=16cm]{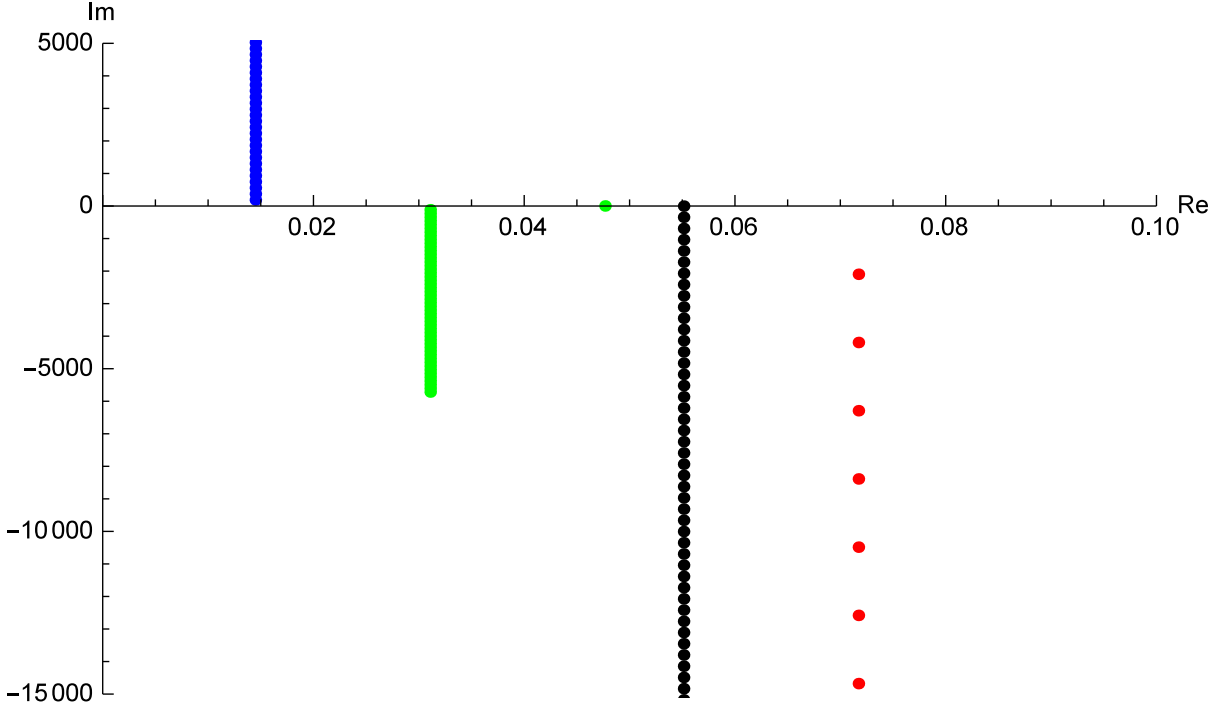}
	\caption{The quasinormal modes ($n=0, 1000, 2000, 3000, ... , \text{ and } 50000$) of charged non-extremal dRGT black hole for the parameters, $M=1, Q=0.5, \Lambda =0.02, q=0.2, \zeta =0, l=5, \gamma =0.05, \text{ and } m_s =0.01$ for the all-region QNMs.}
	\label{Graph_all_modes}
\end{figure}

\begin{figure}[h]
	\includegraphics[width=16cm]{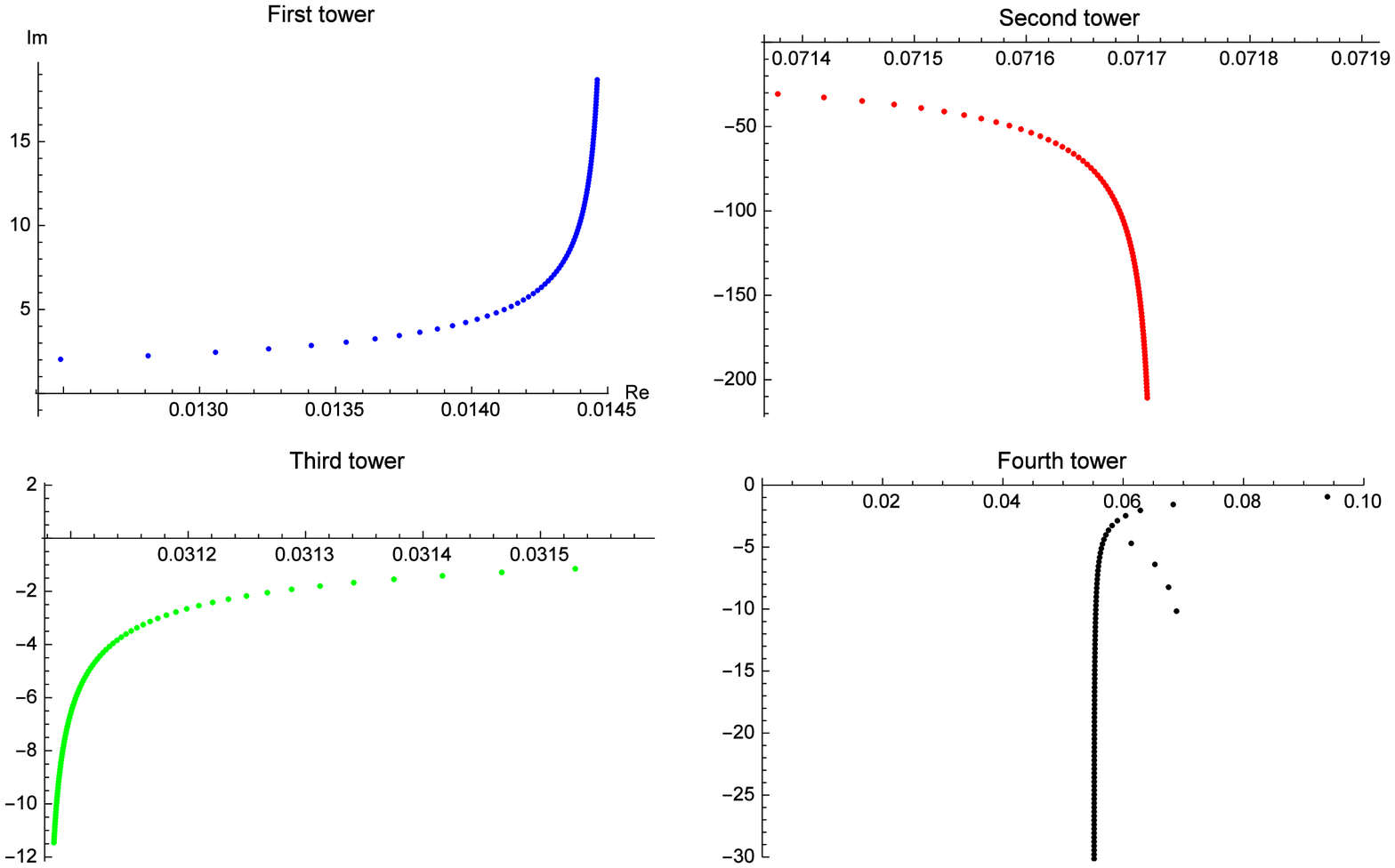}
	\caption{The first 100 quasinormal modes (zoom in) of charged non-extremal dRGT black hole for the parameters, $M=1, Q=0.5, \Lambda =0.02, q=0.2, \zeta =0, l=5, \gamma =0.05, \text{ and } m_s =0.01$ for the all-region QNMs.}
	\label{Graph_zoom_in}
\end{figure}

Set the value of parameters $M=1, Q=0.5, \Lambda =0.02, q=0.2, \zeta =0, l=5, \gamma =0.05, \text{ and } m_s =0.01$ and display the frequencies of the QNMs in Figure \ref{Graph_all_modes} and \ref{Graph_zoom_in}. The four towers of the QNMs are found~(the separated green dot closed to the real axis is the $n=0$ mode lying at $\omega=0.04767+2.14565 i $ of the green tower).

In Figure \ref{Graph_all_modes}, we show the QNMs ($n = 0, 1000, 2000, 3000, ..., \text{ and } 50000$) which are separated into 4 towers. The first tower is represented in the blue dots and it is the unstable modes since the imaginary parts are positive. The remaining 3 towers have negative imaginary parts, they are the decaying quasinormal frequencies represented in the green, black and red towers.

In Figure \ref{Graph_zoom_in} and Table~\ref{convergent_value_all_region_modes}, we show that, as the mode number increases, the real parts of each tower of modes converge to some constant values. Moreover, the gaps in the imaginary parts between modes are also converging to constant values as presented in Table \ref{convergent_value_all_region_modes}. For quasinormal modes (represented in green, black, and red towers), the magnitude of the gap in the imaginary parts increases as the real part of QNMs increases.

\begin{table}[t]
	\centering
	\begin{tabular}{|c|c|c|}
		\hline
		Modes $n \to \infty $ &~~Real parts~~&~~The gap in imaginary parts~~\\
		\hline
		~~The first tower (unstable)~~ & 0.01449 & 0.1862 \\
		The second tower & 0.03108 & -0.1143\\
		The third tower & 0.05512 & -0.3450\\
		The fourth tower & 0.07172 & -2.097\\
		\hline
	\end{tabular}
	\caption{The near-horizon quasinormal modes that can reach far region for $M=1, Q=0.5, \Lambda =0.02, q=0.2, \zeta =0, l=5, \gamma =0.05, \text{ and } m_s =0.01$ in the large $n$ limit.}
	\label{convergent_value_all_region_modes}
\end{table}
Finally, we compare these 4 towers of QNMs with the values obtained by AIM and WKB in the usual coordinates, e.g. as in Ref.~\cite{Burikham:2017gdm}, and found that they are not equal.  They are the near-horizon modes that can reach the far region of the linearly approximated metric background.

\subsection{Exact solution of the QNMs near the event and cosmological horizons in the Rindler coordinate}\label{subsection_Rindler}

From the metric in Eq.~(\ref{metric_gen}), we use the approximation $f(r) = f(r_a) + f'(r_a)(r-r_a) + \mathcal{O}(r-r_a)^2$ to rewrite the metric as
\begin{equation}
ds^2 = -2\kappa_a (r-r_a)dt^2 + \frac{1}{2\kappa_a(r-r_a)}dr^2 + r_a^2d\Omega^2.
\end{equation}
With the definition of the Rindler coordinate $\displaystyle{dx^2 = \frac{dr^2}{2\kappa_a (r-r_a)}}$, the metric in the Rindler spacetime is
\begin{equation}
ds^2 = - \kappa_a^2 x^2 dt^2 + dx^2 + r_a^2 d\Omega^2.
\end{equation}
The Klein-Gordon equation in the Rindler coordinate takes the form
\begin{equation}
x \frac{d}{dx} \bigg( x \frac{d R(x)}{dx} \bigg) + \bigg( \frac{1}{\kappa_a^2}\bigg(\omega - \frac{qQ}{r_a} \bigg)^2 - \bigg(\frac{l(l+1)}{r_a^2} + m_s^2\bigg)x^2 \bigg)R(x) = 0.
\label{KGRindler}
\end{equation}
The general form of solutions can be written as
\be
R(x)=C_{1} J_{\nu}\left(-\frac{i x \sqrt{m^2 r_{a}^2+l(l+1)}}{r_{a}}\right)+C_{2} Y_{\nu}\left(-\frac{i x \sqrt{m^2 r_{a}^2+l(l+1)}}{r_{a}}\right),
\ee
where  
\be
\nu = -\frac{i (r_{a} \omega-q Q)}{\kappa_{a} r_{a}}
\ee
and $J_{\nu}, Y_{\nu}$ are the Bessel function of the first and second kind respectively.   The QNMs for the case when $\nu$ is nonzero integer and half-integer~(half odd-integer) are
\begin{equation}
	\omega = \frac{qQ}{r_a} + i|\kappa_a|\frac{n-1}{2},~~~~~\text{for}~~~~~n=0,-1,-2,...
\end{equation}
The QNMs with negative imaginary parts are the tower of decaying modes.  Comparing to Eq.~(\ref{anao}), the modes with imaginary parts equal to half-integer of the surface gravity, e.g. $i\kappa_{a}/2, 3 i\kappa_{a}/2,...$ emerge in the Rindler coordinate.  Numerical results by the modified AIM confirms emergence of these new modes.  

\section{Discussions and Conclusions}\label{secDiscussion}

The near-horizon quasinormal frequencies of the charged scalar field in the black hole spacetime are investigated numerically and analytically. Using AIM, the tower patterns in the QNMs for any generalized black hole parameters are numerically found.  For the near-extremal Schwarzschild and Reissner-Nordstr$\ddot{{\rm o}}$m dRGT cases, the QNMs near the event horizon of the (uncharged)~scalar field are purely imaginary with the gaps between overtones equal to the surface gravity at the horizon.

For the non-extremal Schwarzshild de-Sitter dRGT cases, the quasinormal modes are also purely imaginary. The value of the gaps between each overtone equals to the corresponding surface gravity depending on the observing point. If we observe QNMs near the event~(cosmological) horizon, the steps then equal to the surface gravity at the event~(cosmological) horizon respectively. For the charged black holes interacting with the charged scalar, the real parts of the QNMs are universally given by $qQ/r_h$ (near the event horizon) and $qQ/r_c$ (near the cosmological horizon). The gaps between overtones are equal to the corresponding surface gravity. 

Finally, we find the analytic solution of the QNMs near the horizons of static spherically symmetric black holes.  The tower pattern of quasinormal frequencies near the horizon is $\omega = \displaystyle{\frac{qQ}{r_a}} + i |\kappa_{a}| n$, where $r_a$ is the corresponding horizon, $\kappa_a$ is the surface gravity at $r_a$, and $n$ is a non-positive integer.  Extending the analysis to the far region of the linearly approximated metric with outgoing-wave boundary condition, four kinds of all-region QNMs are computed as the four roots of a quartic equation.  Each of the four kinds of the QNMs converge to asymptotically constant real parts and constant-spacing imaginary parts.  One tower of the QNMs is unstable while the other three towers are the damping modes.

Ringdown frequency profile of black hole after its formation could contain both the near (event) horizon modes and the all-region~(WKB) mode.  The near-horizon modes usually has vanishing real parts since most black holes do not carry charge but their imaginary parts contain direct information of the surface gravity of the black hole itself.  By more precise data to be collected in the years to come from LIGO/VIRGO and other collaborations, it becomes possible to analyze the damping characteristic of the black hole and detect such modes, gaining very precise information of the black hole from its surface gravity.

\acknowledgments

S.P. is supported by Rachadapisek Sompote Fund for Postdoctoral Fellowship, Chulalongkorn University.  P.B. is supported in part by the Thailand Research Fund (TRF),
Office of Higher Education Commission (OHEC) and Chulalongkorn University under grant RSA6180002.

\end{document}